\documentclass[a4paper,fleqn,usenatbib]{mnras}

\usepackage[T1]{fontenc}
\usepackage{ae,aecompl}

\usepackage{graphicx}	
\usepackage{amsmath}	
\usepackage{amssymb}	

\def\Tau{\mathcal{T}}

\def\gtsima{$\; \buildrel > \over \sim \;$}
\def\simgt{\lower.5ex\hbox{\gtsima}}

\def\songmei2022{https://doi.org/10.48550/arXiv.2212.11034}

\title[SMC star clusters]{The outer structure of old star clusters in the Small Magellanic Cloud}

\author[Andr\'es E. Piatti]{
Andr\'es E. Piatti$^{1,2}$\thanks{E-mail: andres.piatti@fcen.uncu.edu.ar} \\
$^{1}$Instituto Interdisciplinario de Ciencias B\'asicas (ICB), CONICET-UNCUYO, 
Padre J. Contreras 1300, M5502JMA, Mendoza, Argentina\\
$^{2}$Consejo Nacional de Investigaciones Cient\'{\i}ficas y T\'ecnicas, Godoy Cruz 
2290, C1425FQB,  Buenos Aires, Argentina\\
}

\date{Accepted XXX. Received YYY; in original form ZZZ}

\pubyear{2024}

\begin{document}
\label{firstpage}
\pagerange{\pageref{firstpage}--\pageref{lastpage}}
\maketitle

\begin{abstract}
We report results on the internal dynamical evolution of old star clusters
located in the outer regions of the Small Magellanic Cloud (SMC). Because the SMC
has been imprinted with evidence of tidal interaction with the Large Magellanic
Cloud (LMC), we investigated at what extend such an interaction has produced
extra tidal structures or excess of stars beyond the clusters' tidal radii.
For that purpose, we used the Survey of the Magellanic Stellar History
(SMASH) DR2 data sets to build number density radial profiles of
suitable star clusters, and derived their structural and internal dynamics
parameters. The analysed stellar density profiles do not show any evidence of
tidal effects caused by the LMC. On the contrary, the Jacobi volume of the
selected SMC star clusters would seem underfilled, with a clear trend
toward a smaller percentage of underfilled volume as their deprojected distance
 to the SMC centre increases. Moreover, the internal
dynamical evolution of SMC star clusters would seem to be influenced by the
SMC gravitational field, being star clusters located closer to the SMC centre
in a more advanced evolutionary stage. We compared the internal dynamical
evolution of SMC old star clusters with those of LMC and Milky Way
globular clusters, and found that Milky Way globular clusters
have dynamical evolutionary 
paths similar to LMC/SMC old star clusters located closer to their respective galaxy's centres.
Finally, we speculate with the possibility 
that globular clusters belonging to Magellanic Clouds like-mass galaxies
have lived  a couple of times their median relaxation times.
\end{abstract} 




\section{Introduction}

The gravitational field of a galaxy causes that star clusters lose 
stars by tidal heating, from which stellar tidal tails can form. In 
the Milky Way, tidal tails of globular clusters have been searched 
from different observational campaigns. Recently, \citet{zhangetal2022}
compiled an updated sample of globular clusters with detected tidal
tails, along with those showing extended envelopes or without any
signature of extra-tidal structure. From nearly 50 globular
clusters with studies of their outermost regions, more than a half
of them have detected tidal tails. According to \citet{grondinetal2024},
who produced a catalogue of mock extra tidal stars of 159 globular clusters,
tidal tails are a common phenomenon among them, irrespective of whether
they belong to the bulge, disc or halo of the Milky Way.

The effect of the Milky Way gravitational field over the globular clusters
also shapes their internal structures, and hence the pace of their
internal dynamical evolution \citep{gnedinetal1999,fz01,gielesetal2008,webbetal2013,webbetal2014,brockampetal2014,alessandrinietal2014}. \citet{piattietal2019b} showed that
globular clusters' radii increase in general as the Milky Way
potential weakens, with the core and tidal radii being those which
increase at the slowest and fastest rate, respectively. They interpreted
this result as the innermost regions of a globular cluster are less sensitive 
to tidal forces than the outermost ones at a fixed Galactocentric distance. Likewise, 
the Milky Way gravitational field would seem to have differentially accelerated 
the internal dynamical evolution of globular clusters, with those toward the 
bulge appearing dynamically older. Furthermore, the $A^+$ index, which measures
the level of radial segregation of blue straggler stars in old star clusters,
commonly known as the dynamical clock for the long-term
internal dynamical evolution \citep{alessandrinietal2016}, shows
a non negligible dependence on the strength of the host galaxy gravitational
potential \citep{piatti2020c}.

In the Large Magellanic Cloud (LMC), a galaxy nearly an order of magnitude less 
massive than the Milky Way, the effect of tides on its globular clusters has also 
been investigated. \citet{pm2018} traced the stellar density and surface brightness 
radial profiles of almost all globular clusters and derived their structural 
parameters. They found that innermost globular clusters contain an excess of stars
in their outermost regions with respect to the stellar density expected from a King 
profile. Such an excess of stars, not seen in the outermost globular clusters, shows
a clear dependence with the globular cluster's position in the galaxy. The size of the
outermost globular clusters resulted also larger. Although the masses
of all the globular clusters are  commensurate, the outermost regions of those
located in the innermost LMC regions appear to have dynamically evolved more
quickly. These outcomes confirm that globular clusters have mostly been experiencing the 
influence of the LMC gravitational field at their respective mean distances from the LMC 
centre.

The Small Magellanic Cloud (SMC) is a galaxy even an order of magnitude less 
massive than the LMC and, as far as we are aware, there has not been any 
studies on the intensity of the effect of galactic tides (if any) on its old 
globular clusters. Moreover, the SMC has been shaped from the interaction with
the LMC, witnessed by the formation of the Magellanic Bridge, and the presence
of tidally perturbed regions, namely, the Southern bridge, the SMC wing, the West
halo, the Counter bridge, among others 
\citep[][and references therein]{detal14,oliveiraetal2023}.
Therefore, old globular
clusters populating the outermost SMC regions could have also been reached by the
LMC gravitational field. Precisely, the aim of this work consists in analysing 
the structural and internal dynamical evolution parameters of SMC globular clusters 
in order to assess at what extent tidal effects have taken place. In Section~2, we describe
the selection of the globular cluster sample and the construction of their
star number density radial profiles. Section~3 deals with the estimates
of the structural and astrophysical properties for the globular cluster sample, 
and their comparison with published values, while in Section~4 we analyse
the relationship between different resulting parameters in the context of
galactic tidal effects. Section~5 summarizes the main conclusions of this work.

\section{Data handling}

\begin{figure}
\includegraphics[width=\columnwidth]{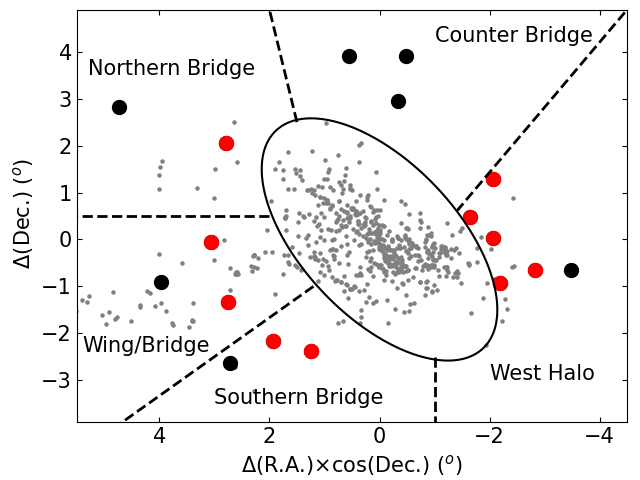}
\caption{Distribution of SMC star clusters in the plane of the sky 
\citep[grey dots,][]{bicaetal2020}. Large filled circles represent 
star clusters older than 4 Gyr; the red ones being those analysed 
in this work (see text for details).}
\label{fig1}
\end{figure}

The SMC has a nearly 1:2:4 3D shape,  with the declination, right ascension and 
line-of-sight being the three axes \citep{cetal01,ripepietal2017}. Hence, 
\citet{petal07} proposed an elliptical framework with position angle 54$\degr$ and 
semi-major to semi-minor axes ratio of 1/2 to represent more meaningfully the 
distribution of star clusters projected on the sky. We here adopted an ellipse with 
a semi-major axis of 2$\degr$ to embrace the SMC main body, and selected star clusters 
older than 4 Gyr \citep{bicaetal2020} located outside that ellipse. Figure~\ref{fig1} 
depicts the distribution of the SMC star cluster population and highlights the selected 
ones.

In order to build the star number density profiles reaching well beyond the outskirts 
of the star clusters, we decided to use the publicly available Survey of the Magellanic
Stellar History (SMASH) DR2 data sets \citep{nideveretal2021}. We accessed the photometry
through the portal of the Astro Data Lab\footnote{https://datalab.noirlab.edu/smash/smash.php}, 
which is part of the Community Science and Data Centre of NSF’s National Optical Infrared 
Astronomy Research Laboratory, to retrieve R.A and Dec. coordinates, PSF $g,i$ magnitudes 
and their respective errors, interstellar reddening $E(B-V)$ and $\chi^2$ and {\sc sharpness} 
parameters of stellar sources located inside a radius of 18$\arcmin$ from the star clusters' 
centres. For seven selected star clusters (BS~196, ESO~51SC-09, Lindsay~1, 32, 38, 112, and 
113) we did not gather any information from the SMASH DR2 database.
The retrieved data sets consist of sources with 0.2 $\le$ {\sc sharpness} $\le$ 1.0  and
$\chi^2$ $<$ 0.5, so that we excluded bad pixels, cosmic rays, galaxies, and unrecognized 
double stars.

Depending on the cluster stellar density, the foreground Milky Way star field and the
SMC composite star field populations, the limiting magnitude and the quality of the PSF $g,i$ 
photometry along the line-of-sight, the colour-magnitude diagrams (CMDs) of the selected star 
clusters show Main Sequences with different extensions. The Main Sequence of 47 Tuc is
also clearly visible and superimposed to some of the star cluster CMDs. For these reasons, 
although Main Sequence stars are generally more numerous than Red Giant Branch (RGB) stars, 
and hence preferred for building stellar density radial profiles \citep{carballobelloetal2012}, 
we decided to rely the present analysis on RGB stars. For each selected SMC star cluster we employed 
different sections of their RGBs with the aim of minimizing the field star contamination and 
maximizing the presence of cluster stars. By restricting the present analysis to cluster RGB stars 
we also deal with 
a photometry completeness much higher than 50$\%$, which for the DECam imager 
\citep{flaugheretal2015} used by SMASH corresponds to $g,i$ $\approx$ 23.3 mag \citep{piatti2021c}.
The final selected star cluster sample includes
Bruck~168, HW~5, 66, 79, Kron~1, 3, Lindsay~2, 109, 110, and NGC~121, which are drawn with red filled circles
in Figure~\ref{fig1}.

\begin{table*}
\caption{Structural and astrophysical properties of SMC star clusters.}
\label{tab1}
\begin{tabular}{lccccc}\hline\hline
Property                &  Bruck~168         & HW~5          & HW~66         & HW~79         & Kron~1         \\\hline
$PA$ ($\degr$)          &  53.7          & 286.4         & 152.5         & 138.2         & 247.2               \\
$r_c$ ($\arcmin$)       &  0.15$\pm$0.03 & 0.35$\pm$0.05 & 0.60$\pm$0.10 & 0.60$\pm$0.10 & 0.70$\pm$0.10  \\
$r_h$ ($\arcmin$)       &  0.30$\pm$0.03 & 0.85$\pm$0.05 & 1.20$\pm$0.10 & 1.20$\pm$0.10 & 1.40$\pm$0.10 \\
$r_t$ ($\arcmin$)       &  1.30$\pm$0.30 & 7.00$\pm$1.50 & 6.00$\pm$1.00 & 6.00$\pm$1.00 & 6.00$\pm$1.00  \\
$r_{cls}$ ($\arcmin$)   &  0.90$\pm$0.20 & 2.50$\pm$0.50 & 5.10$\pm$0.50 & 4.90$\pm$0.50 & 1.50$\pm$0.50  \\
$r_{proj}$ ($\degr$)    &  3.47          & 1.67          & 2.76          & 3.10          & 2.49              \\
$d$ (kpc)               &  61.90$\pm$2.10& 62.00$\pm$4.00& 66.07$\pm$1.52& 60.26$\pm$1.35& 60.26$\pm$2.70\\
$\gamma$                &  4.0$\pm$0.5   & 2.2$\pm$0.1   & 2.3$\pm$0.2   & 3.3$\pm$0.2   & 4.0$\pm$1.0    \\
log($\mathcal{M}_{cls}$/$M_{\odot}$)&3.54&  3.80         &  5.00         & 5.10          & 4.40            \\
$t_r$ (Gyr)             &  1.0$\pm$0.1   & 2.3$\pm$0.2   & 12.3$\pm$1.5  & 11.8$\pm$1.5  & 8.2$\pm$0.9    \\
$E(B-V)$ (mag)          &  0.06$\pm$0.02 & 0.09$\pm$0.03 & 0.05$\pm$0.01 & 0.06$\pm$0.01 & 0.03$\pm$0.01  \\
age (Gyr)               &  6.6$\pm$0.9   & 4.9$\pm$0.4   &  4.0$\pm$0.5  & 5.0$\pm$0.5   & 6.8$\pm$1.0  \\
${\rm [Fe/H]}$ (dex)    &  -1.08$\pm$0.09& -1.30$\pm$0.10& -1.20$\pm$0.10& -1.10$\pm$0.10& -1.08$\pm$0.04\\\hline

Property                &  Kron~3        & Lindsay~2    & Lindsay~109   & Lindsay~110   & NGC~121 \\\hline
$PA$ ($\degr$)          &  270.9         &  256.9       &    115.87     &   90.8          &  302.1        \\
$r_c$ ($\arcmin$)       &  0.50$\pm$0.05 & 0.35$\pm$0.05& 0.30$\pm$0.05 & 0.60$\pm$0.05 & 0.20$\pm$0.02 \\
$r_h$ ($\arcmin$)       &  1.10$\pm$0.10 & 0.60$\pm$0.05& 0.55$\pm$0.05 & 1.05$\pm$0.05 & 0.60$\pm$0.05 \\
$r_t$ ($\arcmin$)       &  6.00$\pm$1.00 & 2.00$\pm$0.50& 2.00$\pm$0.50 & 2.10$\pm$0.10 & 6.00$\pm$0.50 \\
$r_{cls}$ ($\arcmin$)   &  2.50$\pm$0.30 & 1.80$\pm$0.30& 1.10$\pm$0.20 & 1.70$\pm$0.10 & 4.60$\pm$0.50 \\
$r_{proj}$ ($\degr$)    &  2.06          &   2.91       &  3.07         & 3.08          & 2.31          \\
$d$ (kpc)               &  61.09$\pm$2.00&56.60$\pm$1.80& 62.50$\pm$2.00& 57.50$\pm$2.30& 64.60$\pm$0.40 \\
$\gamma$                &  3.3$\pm$0.2   &  4.0$\pm$0.5 &  4.0$\pm$0.50 & 6.0$\pm$1.0   & 2.4$\pm$0.1 \\
log($\mathcal{M}_{cls}$/$M_{\odot}$)&4.80&   ---        & ---           & 4.20          & 5.60  \\
$t_r$ (Gyr)             &  8.1$\pm$1.0   &   ---        & ---           & 7.2$\pm$0.8   & 7.3$\pm$0.9 \\
$E(B-V)$ (mag)          &  0.02$\pm$0.01 & 0.10$\pm$0.03&  0.07$\pm$0.02& 0.05$\pm$0.02 & 0.04$\pm$0.01 \\
age (Gyr)               &  5.6$\pm$1.0   & 4.0$\pm$0.2  &  4.5$\pm$0.5  & 6.3$\pm$1.0   & 9.7$\pm$1.0 \\
${\rm [Fe/H]}$ (dex)    &  -0.90$\pm$0.10&-1.28$\pm$0.09& -1.10$\pm$0.15 &-1.03$\pm$0.05& -1.18$\pm$0.02 \\\hline

\end{tabular}
\end{table*}

\begin{figure}
\includegraphics[width=\columnwidth]{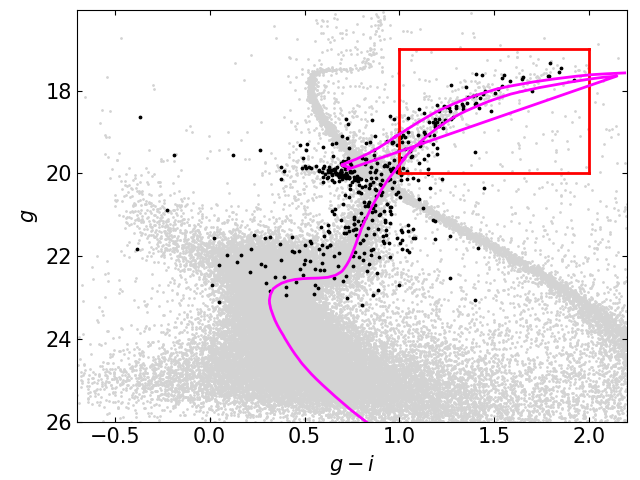}
\caption{Colour-magnitude diagram of stars measured by SMASH in the field of NGC~121 (grey dots).
The black points correspond to stars distributed inside 3$\times$$r_c$ (see Table~\ref{tab1})
from the cluster's centre. An isochrone for the cluster's parameters is overplotted.
The red rectangle shows the cluster RGB section employed to built the star number density
radial profile (see text for details).}
\label{fig2}
\end{figure}

Figure~\ref{fig2} shows the CMD built with all the stars measured by SMASH located in the 
field of NGC~121, represented with grey points. As can be seen, the predominantly old SMC composite
field star population and the long Main Sequence of 47 Tuc are clearly visible. With the
aim of uncovering the cluster's RGB, we painted black stars distributed inside a circle centred
on the cluster and with a radius 3 times the cluster's core radius ($r_c$, see Table~\ref{tab1}).
In practice, we started by using circles with radii between 1$\arcmin$ and 2$\arcmin$, which
resulted to be nearly 3$\times$$r_c$ (see below the estimation of $r_c$). Then, we inspected the 
NGC~121's CMD and decided on the RGB section to use for constructing its star number density
radial profile. We drew in Figure~\ref{fig2} a red rectangle embracing the selected
RGB section. We proceeded similarly with the remaining selected star clusters. For
completeness purposes, we provide in the Appendix the corresponding clusters' CMDs.

The star number density radial profiles were built by counting the number of stars distributed 
inside boxes of 0.02$\arcmin$ up to 1.00$\arcmin$ per side, increasing by steps of 0.02$\arcmin$ 
a side. We then built mean radial profiles with their uncertainties by averaging the resulting
individual ones. As a matter of fact, the radial profiles built from star counts performed using
smaller boxes result smoother toward the innermost clusters' regions, while those obtained from 
larger boxes, trace better the ample clusters' surrounding fields. We fitted with a
constant value the background level, using points located from 10$\arcmin$ up to 18$\arcmin$
from the clusters' centrers. Although the mean background value is different for each cluster field
(from 1.3 up to 6 stars/arcmin$^2$), its dispersion resulted to be between 3$\%$ and 10$\%$ of the 
mean background level. The intersection between the observed radial profile and the fitted background
level was used to define the cluster radius ($r_{cls}$, see Table\ref{tab1}). We then subtracted 
from the observed radial profile the mean value of the background stellar density derived above. 

Figure~\ref{fig3} shows the observed and background subtracted radial profiles obtained for NGC~121 
drawn with open and black filled circles, respectively. The 
error bars represent the dispersion of the observed radial profile and that of the background level 
added in quadrature. \citet{getal09} traced the cluster radial profile from $HST$ data, which we 
superimposed to Figure~\ref{fig3} with open squares for completeness purposes. As can be
seen, although the present radial profile is much more extended, both agree very well in the innermost
overlapping cluster's region, which implies that the present approach of using RGB stars
does not suffer from star count incompleteness. We finally fitted three different profile models 
to the normalized background-corrected radial profiles with the aim of deriving the clusters' core 
($r_c$), half-light ($r_h$) and tidal ($r_t$) radii, respectively. The \citet{king62}'s profile model 
was used to derive $r_c$ and $r_t$; the \citet{plummer11}'s model was applied to obtain $r_h$ from 
the relation $r_h$ $\sim$ 1.3$\times$a; and the \citet{eff87}'s model provided $r_c$ and $\gamma$ 
($\gamma$ is the power-law slope at large radii). The resulting values of these structural parameters 
are listed in Table~\ref{tab1}, while Figure~\ref{fig3} illustrates how they reproduce the NGC~121 
radial profile. The Appendix includes figures similar to Figure~\ref{fig3} for the remaining selected 
star clusters. Whenever very few or no RGB stars were found in the selected clusters' regions,
we focused on the cluster's sub-giant branch regions instead.

\subsection{SMC star cluster properties}

As far as we are aware, the SMC contains 17 star clusters older than
4 Gyr located in its outer disc, namely, those highlighted with large red and 
black circles in Figure~\ref{fig1}. We have thoroughly searched the available 
literature looking for reliable structural parameters for these 17 star
clusters. Although there have been some previous studies on the structural parameters 
of a large sample of SMC star clusters, among them, those by \citet{hz2006}, 
\citet{santosetal2020} and \citet{gattoetal2021}, only \citet{mg2003b} and \citet{getal09} 
derived core and tidal radii and other astrophysical properties for NGC~121. 
This implies that the present resulting structural and internal dynamics parameters 
for other nine star clusters (Bruck~168, HW~5, 66, 79, Kron~1, 3, Lindsay~2, 109, and 110) 
represent a significant increase in the number of outermost old SMC clusters
with  estimates of these properties. At the same time, this outcome points to
the need of further investigations of the structure and internal dynamics of the 
remaining 10 old SMC populating the galaxy outer disc.

\begin{figure}
\includegraphics[width=\columnwidth]{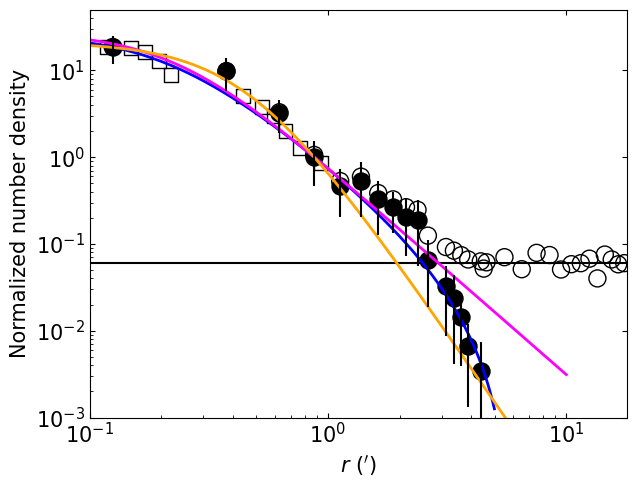}
\caption{Normalized observed and background-corrected star number density radial profiles of NGC~121 
drawn with open and black filled circles, respectively, with uncertainties represented by error bars. 
Radial profiles are normalized to the star number density at $r$= 0.875$\arcmin$.
The horizontal line represents the adopted mean background level. The open squares correspond to the 
radial profile built by \citet{getal09}. Blue, orange, and magenta solid lines are the
best-fitted \citet{king62}'s, \citet{plummer11}'s, and \citet{eff87}'s models, respectively.}
\label{fig3}
\end{figure}

Since we are interested in comparing the different clusters'
linear radii and relaxation times of the selected SMC star clusters with those of old 
globular clusters in the LMC and in the Milky Way, we gathered from the literature age 
and distance estimates for the studied cluster sample. In the case of HW~66 and 79, we derived these 
astrophysical properties using the models of \citet[][PARSEC v1.2S\footnote{http://stev.oapd.inaf.it/cgi-bin/cmd}]{betal12} 
for the SMASH photometric system and routines of the Automated Stellar 
Cluster Analysis code \citep[ASteCA,][]{pvp15}. For the sake of the reader, 
the isochrones for the adopted clusters' parameters are superimposed to the respective CMDs 
(see Figure~\ref{fig3}, and A.1-A.6). As can be seen, the isochrones for the the adopted 
star clusters' properties satisfactorily match the observed clusters' CMDs.

The linear values (in parsecs) of $r_c$, $r_h$, $r_t$ and $r_{cls}$ were obtained
using the expression 2.9 10$^{-4}$ $d$ $r_{c,h,t}$, where $d$ is the star
cluster's heliocentric distance (see Table~\ref{tab1}). We also derived the
star cluster's deprojected distance (in kpc) using the relation: \\

$r_{deproj}$ = ($d_{smc}$$^2$ +  $d^2$ - 2 $d_{smc}$ $d$ cos($r_{proj}$))$^{1/2}$, \\

\noindent where $d_{smc}$ is the heliocentric distance of the SMC centre
\citep[62.5$\pm$0.8 kpc, ][]{graczyketal2020}.

We computed the star cluster masses ($M_{cls}$) using the relationships obtained by 
\citet[][equation 4]{metal14} as follows:\\

log($\mathcal{M}_{cls}$ /$M_{\odot}$) = $a$ + $b$ log(age /yr) - 0.4 ($M_i - M_{i_{\odot}}$),\\

\noindent where $M_i$ is the integrated absolute magnitude in the filter $i$ = $B,V$ 
of the Johnson photometric system ($M_{B_{\odot}}$ = 5.49 mag, $M_{V_{\odot}}$ = 4.83 mag), 
and $a$ and $b$ are from Table~2 of \citet{metal14} 
for a representative SMC overall metallicity of $Z$ = 0.004 \citep{pg13}. The absolute
integrated magnitude were computed from the observed ones as follows:\\

$M_i$ = $i$ - $A_i$ - 5 log($d$/10),\\

\noindent where $i$ is the observed $B,V$ integrated magnitude, and $A_i$ the
mean interstellar absorption (see Table~\ref{tab1}). Typical uncertainties turned out 
to be $\sigma$(log($\mathcal{M}_{cls}$/$M_{\odot}$)) $\approx$ 0.2.

We calculated half-light relaxation times using the equation of \citet{sh71}:\\

$t_r = (8.9\times 10^5 \mathcal{M}_{cls}^{1/2} r_h^{3/2}) / (\bar{m} log_{10}(0.4\mathcal{M}_{cls}/\bar{m}))$,\\

\begin{figure}
\includegraphics[width=\columnwidth]{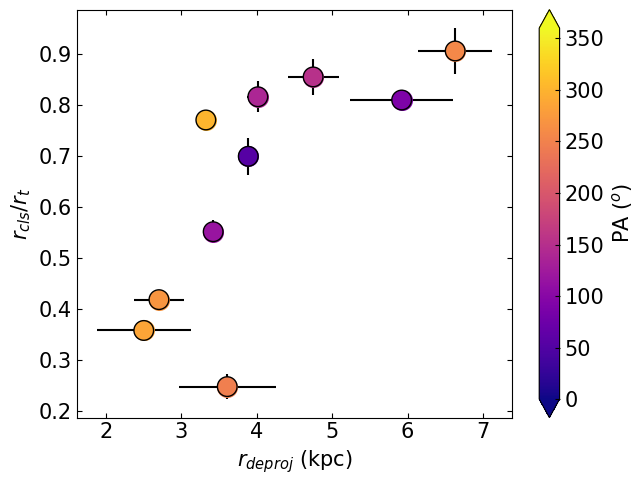}
\caption{Relationship between $r_{cls}/r_t$ and the deprojected
cluster distance to the SMC centre, coloured according to the
cluster position angle. Error bars are also drawn.}
\label{fig4}
\end{figure}

\noindent where $\bar{m}$ is the average mass of the cluster stars. For simplicity we assumed a 
constant average stellar mass of 0.5 M$_{\odot}$, which corresponds to the average between the
smallest (0.09$M_{\odot}$) and the largest one (0.92$M_{\odot}$) in the theoretical
isochrones of \citet{betal12} for the selected SMC star clusters.
The uncertainties of the above computed quantities were obtained by performing a thousand
Monte Carlo experiments from which we calculated the standard deviation.

For Bruck~168, HW~5 and Lindsay~2 we adopted the redenning, the distance, the age and the 
metallicity ([Fe/H]) derived by \citet{diasetal2021} and \citet{saroonetal2023}, respectively, 
both from a relative deep $B,V$ photometry obtained with the 4.1 m 
Southern Astronomical Research (SOAR) telescope, which has a Ground Layer Adaptive Optics (GLAO) 
module installed in the SOAR Adaptive Optics Module Imager \citep[SAMI ,][]{tokovininetal2016}
(see Table~\ref{tab1}). The integrated magnitudes for HW~5 were taken from 
\citet[][$B$=15.44 mag, $V$= 14.64 mag.]{retal05}, while the Bruck~188's mass was taken from
\citet{santosetal2020}. We fitted theoretical isochrones to the
CMDs of HW~66 and 79, and adopted the mass obtained by \citet{kontizasetal1986}.
For Kron~1 and 3, we adopted the redenning, the distance, the age and the metallicity
obtained by \citet{miloneetal2023} from HST images. They masses come from integrated photometry
provided by \citet{alcaino1978}: $B$ = 13.9 mag for Kron~1, and $B$=12.74 mag, $V$ = 12.05 mag
for Kron~3. The resulting mass for Kron~1 is in excellent agreement with that obtained
by \citet{gattoetal2021} using images obtained at the ESO VLT Survey Telescope (VST), equipped
with the OmegaCAM mosaic camera, and the $g,r,i$ passbands. For Lindsay~110 we used the
redenning, the distance, the age and the metal content derived by \citet{petal15a}, and the
cluster mass from \citet{kontizasetal1986}. NGC~121 is by far the selected SMC star cluster 
with more estimates of its astrophysical properties. We adopted the redenning, the distance and
the age from \citet{miloneetal2023}, and the metallicity from \citet{mucciarellietal2023}, who
used high-dispersion spectra obtained with FLAMES at the ESO VLT. We computed the cluster
mass from its integrated magnitudes, $B$= 11.2 mag, $V$=11.24 mag \citep{alcaino1978}, and
the resulting value shows an excellent agreement with that derived by \citet{mg2003b}. They
also estimated $r_c$ for NGC~121, which shows an excellent agreement with both the \citet{getal09}
and the obtained in this work (see Table~\ref{tab1}).

\begin{figure}
\includegraphics[width=\columnwidth]{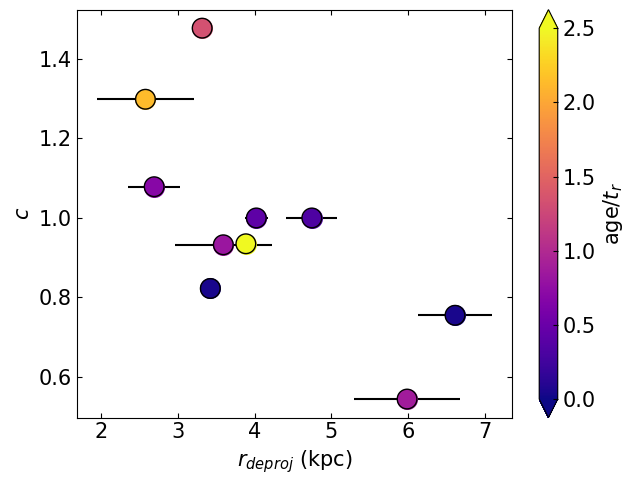}
\caption{Relationship between $c$ and the deprojected
cluster distance to the SMC centre, coloured according to the
age/$t_r$ ratio. Error bars are also drawn.}
\label{fig5}
\end{figure}

\section{Data analysis and discussion}

In this Section, we analyse the relationship between the different SMC clusters' properties
listed in Table~\ref{tab1}, with the aim of searching for differential tidal effects within
the SMC, a behaviour already documented for Milky Way and LMC globular clusters. At the same 
time, we compare the clusters' properties in these three galaxies, which distinguish 
themselves for having different total masses, with a mass relation $\sim$ 100:10:1 for the 
Milky Way, the LMC and the SMC, respectively. Therefore, cluster structural and internal 
dynamics parameters are analysed along a wide galaxy mass baseline. As for LMC globular 
clusters, we used the parameter values published by \citet{pm2018}, while for Milky Way 
globular clusters we made use of the compilation in \citet{piattietal2019b}.

\begin{figure}
\includegraphics[width=\columnwidth]{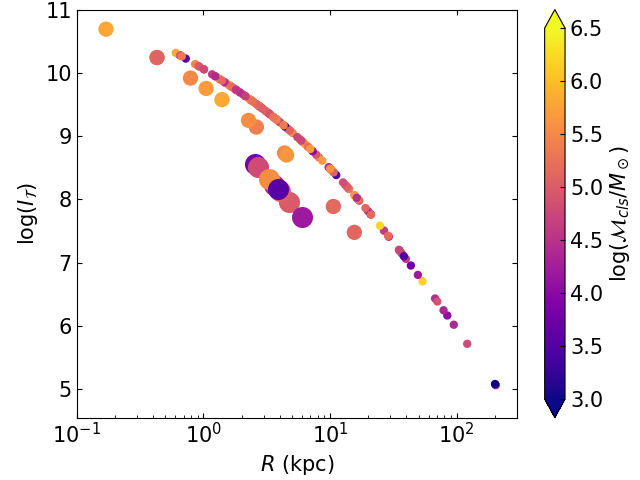}
\caption{Relationship between the distance $R$ to the galaxy's centre
and the tidal index for Milky Way (small circle), LMC
(medium-size circle), and SMC (large circle) globular clusters, coloured 
according to the globular clusters' masses, respectively.}
\label{fig6}
\end{figure}

Figure~\ref{fig4} shows the trend of the ratio between the cluster radius ($r_{cls}$) and 
the  \citet{king62}'s tidal radius ($r_t$) with the cluster deprojected distance to the SMC 
centre. As can be seen, $r_{cls}/r_t$ is smaller than the unity for all the selected star
clusters, which means that they still have enough room to expand within the Jacobi
volume, without any star escaping from them. These star clusters are tidally underfilled and
show a trend toward being tidally filled as their deprojected distances increase. 
Such a behaviour was predicted theoretically by \citet{hm2010} and \citet{bianchinietal2015},
among others. Star clusters in weaker tidal fields, like those located in the outermost 
regions of the SMC can expand naturally, whereas those immersed in stronger tidal fields 
do not. We have coloured the points in Figure~\ref{fig4} according to the position angle of 
the star clusters, measured from North to East. By looking at Figure~\ref{fig1}, it is 
straightforward to conclude that the farthest star clusters from the SMC centre in 
Figure~\ref{fig4} are those closer to the LMC. Because these star clusters are those most 
expanded, they experience weaker tidal fields than those on the opposite side of the SMC. 
Hence, we interpret this outcome as that the LMC gravitational field is less intense than 
the SMC potential at $\sim$ 6 kpc from the SMC.

The above mentioned expansion of a star cluster under the effect of a weaker external tidal 
field is concurrently accompanied by its internal dynamical evolution. The ratio between
the age and the central relaxation time ($t_r$) is a frequently used index to measure
the internal dynamical equilibrium level within a star cluster; the larger age/$t_r$ the 
more virialysed the star cluster. On the other hand, the parameter $c$=log($r_t/r_c$) 
also measures the internal dynamical evolution from the internal cluster structure.
While a star cluster evolves dynamically, the more massive stars sink toward the cluster centre,
so that $r_c$ decreases while $r_t$ keeps nearly constant, making $c$ larger for clusters
in a more advanced internal evolutionary stage. Figure~\ref{fig5} depicts 
the behaviour of both internal dynamical evolutionary indices as a function of the
star cluster deprojected distances to the SMC centre. As can be seen, there is a
 correlation between $c$ and $r_{deproj}$, with larger $c$ values for star clusters
closer to the SMC centre. The more advanced internal dynamical evolutionary stage of
inner star clusters is confirmed by the correlation between age/$t_r$ and $r_{deproj}$.
This result shows that the SMC gravitational field accelerates the internal evolution
of stars clusters, as also the Milky Way and the LMC do on their own globular
cluster populations (see Introduction).

The studied SMC star clusters have masses in  general smaller than those of LMC 
globular clusters, and both star cluster populations together nearly span the mass range of
Milky Way globular clusters (see Figure~\ref{fig6}). In order to assess whether
similar star clusters (similar masses) undergo different internal dynamical evolutionary 
paths in the presence of different external potential well, we  analysed the variation
of structural and internal dynamics parameters as a function of the tidal strength.
Although the distance of a star cluster to its host galaxy's centre has been used 
to monitor the variation of the strength of tidal effects on a star cluster 
\citep{hm2010, bianchinietal2015,pm2018,piattietal2019b}, we here 
attempt a slightly more straightforward approach. The tidal force strength 
($\Tau$) \citep[see, e.g.][]{lokasetal2011,pfefferetal2018} can be expressed as:\\

$\Tau$  $\propto$ $\mathcal{M}_{gal}$($< R$) / $R^3$ = $I_{\Tau}$,\\

\noindent  where $\mathcal{M}_{gal}$($< R$) is the total galaxy mass
 within the distance $R$ (approximated with a spherical distribution as in \citet{pl2022}). 
 In what follows
we use $I_{\Tau}$ as a tidal index, with $\mathcal{M}_{gal}$
and $R$ given in $M_{\odot}$ and kpc units, respectively. As can be seen in
Figure~\ref{fig6},  globular clusters in different galaxies at the same $R$ distance 
have associated different $I_{\Tau}$ values; the more massive a galaxy, the larger the
tidal index, as expected.

\begin{figure}
\includegraphics[width=\columnwidth]{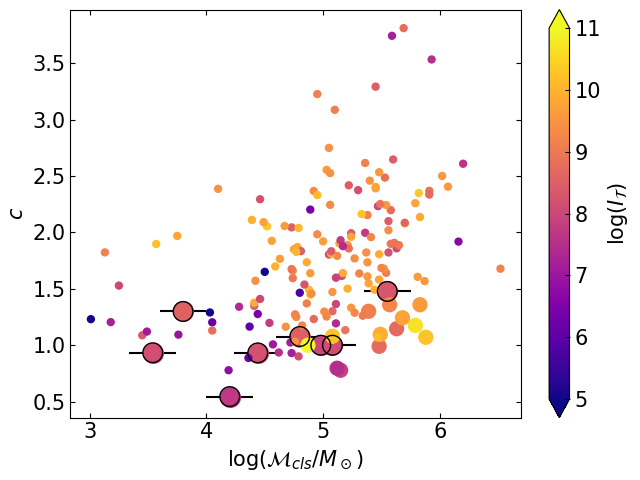}
\caption{Relationship between $c$ and the cluster's mass for Milky Way
(small circle), LMC
(medium-size circle), and SMC (large circle) globular clusters, coloured 
according to their $I_{\Tau}$ values.}
\label{fig7}
\end{figure}

We plotted in Figure~\ref{fig7} 
the $c$ parameter as a function of the cluster's mass for globular clusters in the Milky Way,
in the LMC and in the SMC.   In general, for a fixed cluster's mass, Milky Way globular
clusters are in a more advanced evolutionary stage than LMC and 
SMC old star clusters  (larger $c$ values); a behaviour which would seem to be related 
to their respective tidal force strength. We coloured the globular clusters according to their
tidal index and the arising overall picture leads to confirm that
globular clusters located in the inner Milky Way regions (stronger gravitational field)
have internally evolved faster than those located in the outer Milky Way regions (weaker
gravitational field). In this sense, Figure~\ref{fig6} suggests  that LMC and SMC old
star clusters  have experienced galaxy tidal strengths similar to those of
Milky Way globular clusters located, respectively, at $\sim$ 1.5 and 3.5 times farther from the
galaxy's centre.

\begin{figure}
\includegraphics[width=\columnwidth]{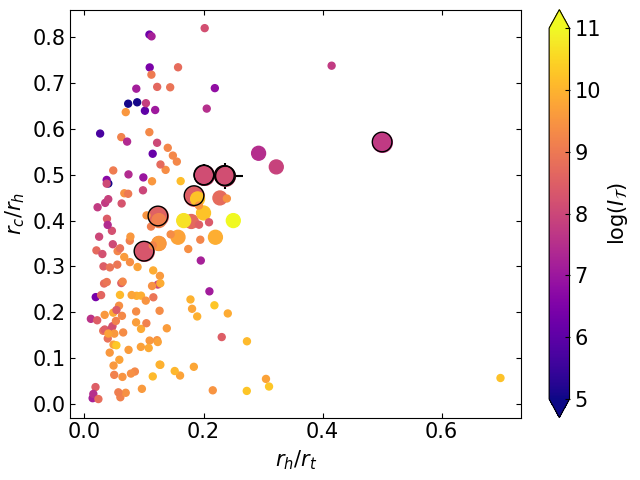}
\caption{Relationship between $r_c/r_h$ and $r_h/r_t$ for Milky Way
(small circle), LMC
(medium-size  circle), and SMC (large circle) globular clusters, coloured according to
their  $I_{\Tau}$ values.}
\label{fig8}
\end{figure}

The internal dynamical evolution of a star cluster is usually monitored in the
$r_c/r_h$ vs $r_h/r_t$ plane, where they move approximately in the top-right-bottom-left 
direction  \citep[][see, e.g., their figure 33.2]{hh03} while relaxing toward a 
core-collapse stage. Figure~\ref{fig8} shows that LMC and SMC old star clusters
evolve differently than the vast majority of Milky Way globular clusters, although
some few of them would seem to share a similar evolutionary path. For Milky Way
globular clusters the variation of $r_h/r_t$ is much smaller than that for
$r_c/r_h$, so that changes in the stellar density profiles are relatively mainly
produced in the innermost clusters' regions. We note that this evolution is
also affected by the effects of the gravitational  field. 
In the case of LMC globular clusters, which have masses and ages
comparable to massive Milky Way globular clusters (log($\mathcal{M}_{cls}/M_{\odot}$) $>$ 5.0), 
their $r_h/r_t$ ratios are larger, which implies earlier internal dynamical
evolutionary stages. The selected SMC star clusters also follow a trend similar
to that of LMC globular clusters, although they are younger and on average less
massive. Lindsay~110 has the largest $r_c/r_h$ and $r_h/r_t$ values
(least evolved), in agreement with their smallest $c$ values (Figure~\ref{fig7}). 

\begin{figure}
\includegraphics[width=\columnwidth]{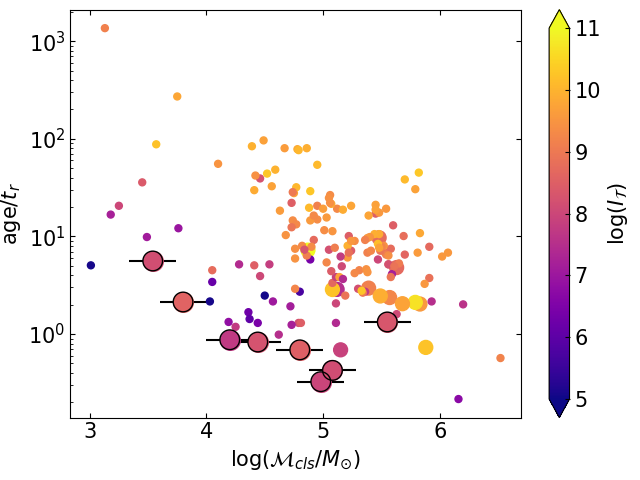}
\caption{Relationship between the cluster's mass and the age/$t_r$ ratio 
for Milky Way (small circle), LMC (medium-size circle), and SMC (large circle) 
globular clusters, coloured 
according to their $I_{\Tau}$ values.}
\label{fig9}
\end{figure}

Finally, Figure~\ref{fig9} illustrates the correlation between the cluster's
mass and the times a star cluster lived its median relaxation time for Milky Way, 
LMC and SMC old star clusters. As can be seen, LMC and SMC star clusters are
the least dynamically relaxed systems, and are comparable from a internal
dynamical evolutionary point of view to outer halo Milky Way globular clusters
($R_{GC}$ $>$ 50 kpc). The LMC is nearly 10 times more massive than the SMC 
and still their respective old cluster populations have dynamically evolved 
similarly (see Figures~\ref{fig7} to \ref{fig9}). Therefore, if we extrapolated
this behaviour to any Magellanic Cloud mass-like galaxies 
and less massive galaxies, we would speculate with
the possibility that their old star clusters had lived less than  a couple of times its median 
relaxation time. Likewise, Milky Way mass-like galaxies and more massive 
galaxies would seem to be the responsible for the faster internal dynamical 
evolution experienced by their old globular clusters.  

\section{Conclusions}

The interaction between both Magellanic Clouds has left imprinted 
some trails on them, such as tidally perturbed regions, formation of 
star clusters, gas outflows, among other observable witnesses.
Being the SMC the least massive of these galaxies, we embarked
in an investigation to probe whether outer old SMC star clusters 
-- as the most suitable candidates -- are reached by the LMC
gravitational field. We sought for the presence of extra-tidal 
structures or excess of stars beyond the clusters' \citet{king62}
tidal radii in old SMC star clusters closer to the LMC, in contrast
to the absence thereof in those star cluster located on the opposite 
side of the SMC.

By deriving the cluster structural parameters from their star number
density radial profiles, we found that not only the selected star clusters
do not exhibit any evidence of tidal effects caused by the LMC,
but also that they are all underfilled, which means that they still
can expand within their respective Jacobi radii. The percentage of
underfilled cluster volume would seem to decrease as the deprojected
distance of the star cluster to the SMC centre increases, an scenario
predicted theoretically for isolated galaxies. Star clusters in the presence of a weaker
external gravitational potential usually expand. We found that the
internal dynamical evolution of the selected star clusters are
influenced by the SMC gravitational field, in the sense that, the closer
a star cluster to the SMC centre (where the gravitational field is 
stronger), the more advanced its internal dynamical evolutionary stage.
This phenomenon was previously observed in old globular clusters of the
LMC and of the Milky Way, whose total galaxy masses are nearly one and two
order of magnitudes larger, respectively.

The effects on the internal dynamical evolution of a population of star 
clusters belonging to the same galaxy, because of the gravitational potential 
of the host galaxy, is also different. It depends on the strength of the
gravitational field at the position of the star clusters. To this respect,
we found that Milky Way globular clusters are in general in a more advanced
evolutionary stage than LMC and SMC old star clusters.  Indeed,
LMC and SMC old star clusters would seem to have experienced tidal strengths
similar to Milky Way globular clusters located, respectively, $\sim$ 1.5 and 3.5 times
farther from the galaxy's centre.
Finally, the times that LMC and SMC
star clusters have lived their median relaxation times are
 in general smaller than Milky Way globular clusters with similar clusters' masses. 

\section*{Acknowledgements}
 We thank the referee for the thorough reading of the manuscript and timely suggestions to improve it.

This research uses services or data provided by the Astro Data Lab at NSF's National 
Optical-Infrared Astronomy Research Laboratory. NSF's OIR Lab is operated by the 
Association of Universities for Research in Astronomy (AURA), Inc. under a cooperative 
agreement with the National Science Foundation.

Data for reproducing the figures and analysis in this work will be available upon request
to the author.

\section{Data availability}

Data used in this work are available upon request to the author.






\begin{appendix}

\section{SMC star clusters' supplementary material}

In this Section we include the resulting CMDs for the selected
star clusters (Figures~A.1-A.9). They show all the stars measured by SMASH in the
clusters' fields, represented by grey dots, and highlight with black
dots stars distributed inside circles with radii equals to 3$\times$$r_c$
centred on the clusters. The red rectangle embraces the cluster RGB
section selected to build the cluster star number density radial
profiles. We have also overplotted the isochrone corresponding to the
adopted cluster's parameters (see Table~\ref{tab1}).

The normalized background-corrected radial profiles of the selected
star clusters are drawn with open circles in Figures~A.10-A.18, with
uncertainties represented by error bars. Blue, orange, and magenta solid lines 
are the best-fitted \citet{king62}'s, \citet{plummer11}'s, and \citet{eff87}'s 
models, respectively (see text for details).

\begin{figure}
\includegraphics[width=\columnwidth]{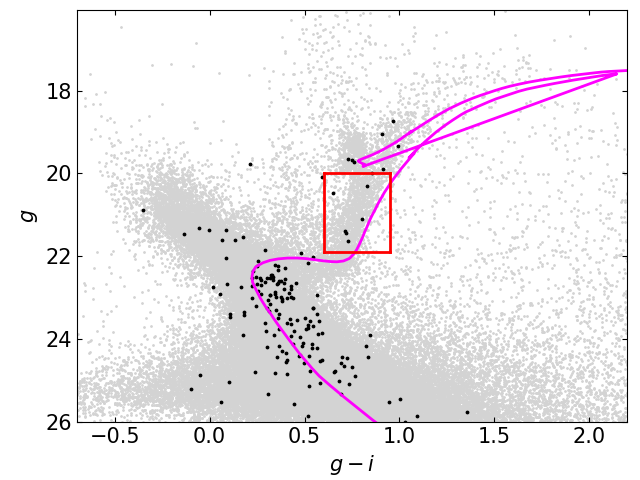}
\caption{Same as Figure~\ref{fig2} for B~168}
\end{figure}

\begin{figure}
\includegraphics[width=\columnwidth]{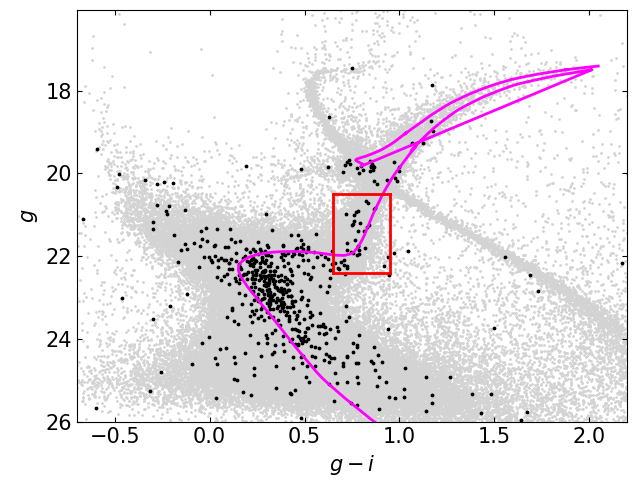}
\caption{Same as Figure~\ref{fig2} for HW~5}
\end{figure}

\begin{figure}
\includegraphics[width=\columnwidth]{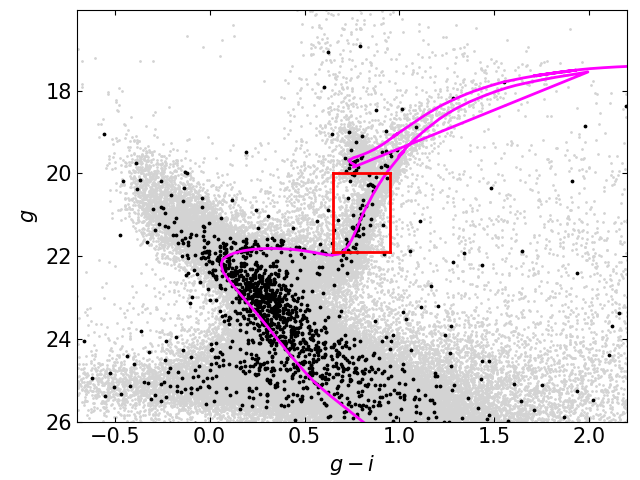}
\caption{Same as Figure~\ref{fig2} for HW~66}
\end{figure}

\begin{figure}
\includegraphics[width=\columnwidth]{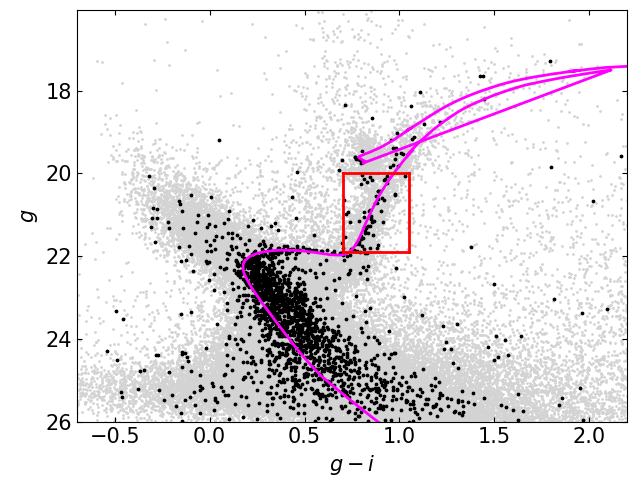}
\caption{Same as Figure~\ref{fig2} for HW~79}
\end{figure}

\begin{figure}
\includegraphics[width=\columnwidth]{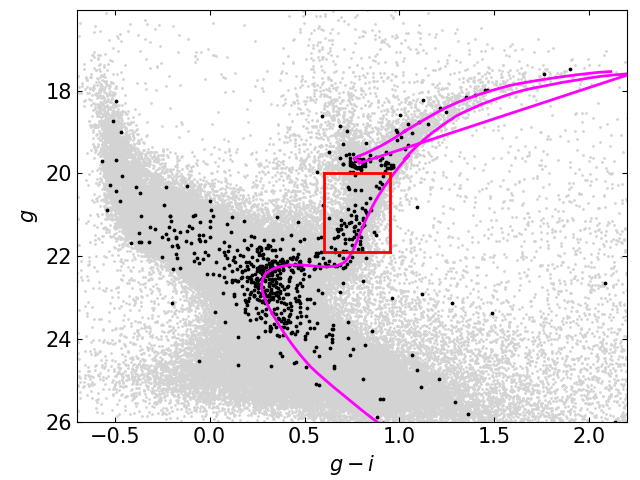}
\caption{Same as Figure~\ref{fig2} for K~1}
\end{figure}

\begin{figure}
\includegraphics[width=\columnwidth]{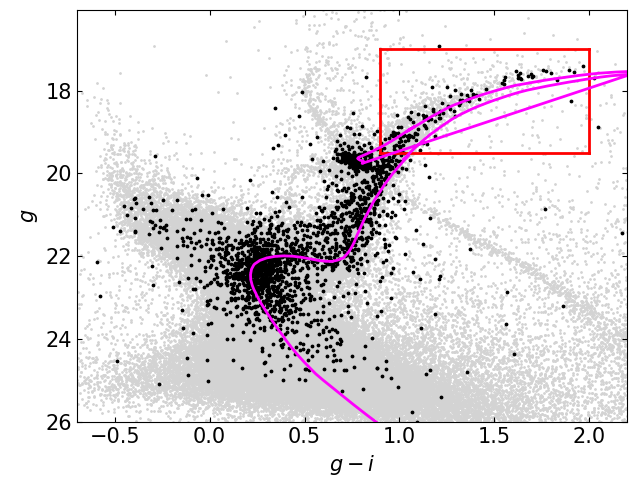}
\caption{Same as Figure~\ref{fig2} for K~3}
\end{figure}

\begin{figure}
\includegraphics[width=\columnwidth]{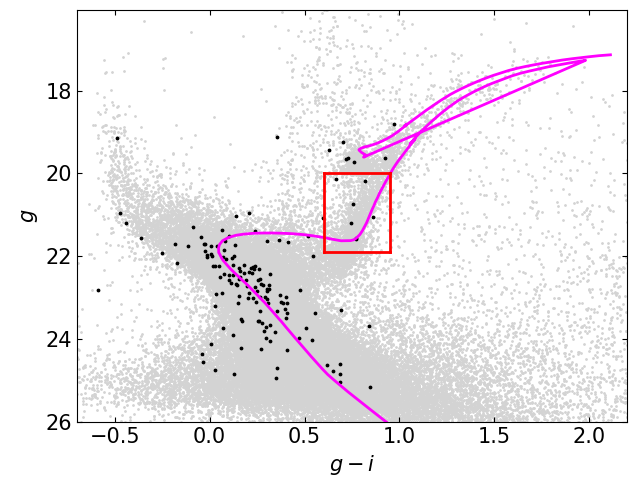}
\caption{Same as Figure~\ref{fig2} for L~2}
\end{figure}

\begin{figure}
\includegraphics[width=\columnwidth]{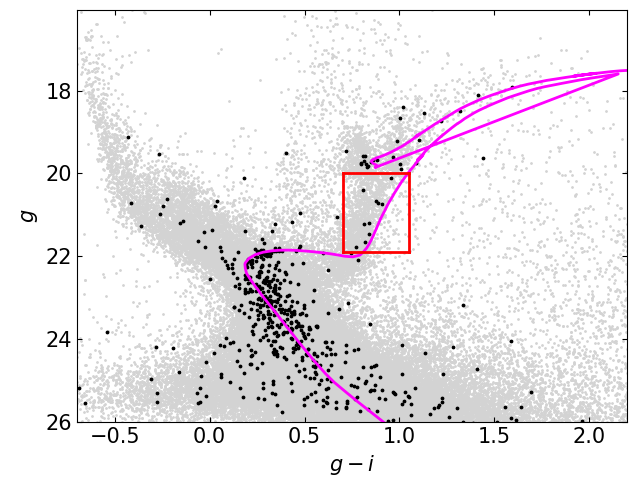}
\caption{Same as Figure~\ref{fig2} for L~109}
\end{figure}

\begin{figure}
\includegraphics[width=\columnwidth]{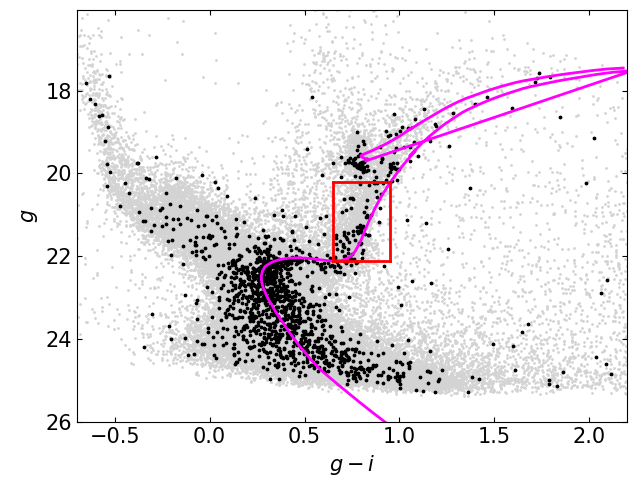}
\caption{Same as Figure~\ref{fig2} for L~110}
\end{figure}

\begin{figure}
\includegraphics[width=\columnwidth]{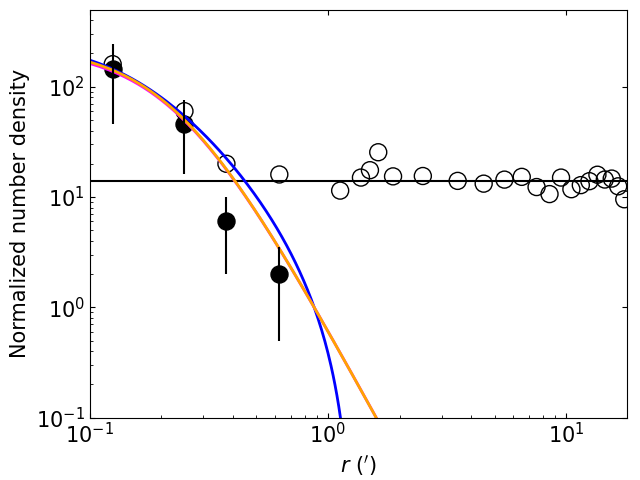}
\caption{Same as Figure~\ref{fig3} for B~168}
\end{figure}

\begin{figure}
\includegraphics[width=\columnwidth]{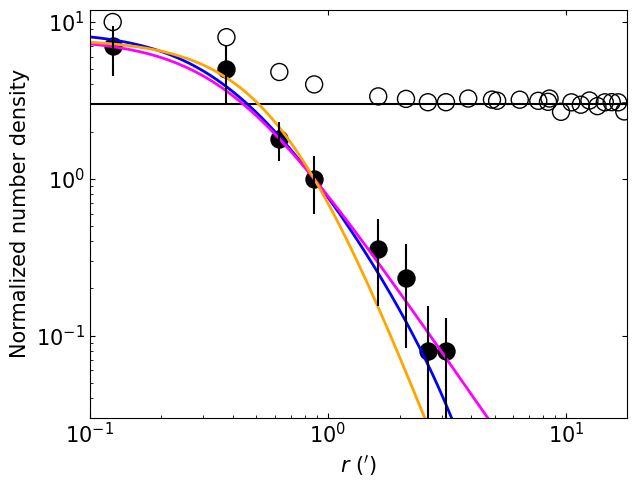}
\caption{Same as Figure~\ref{fig3} for HW~5}
\end{figure}

\begin{figure}
\includegraphics[width=\columnwidth]{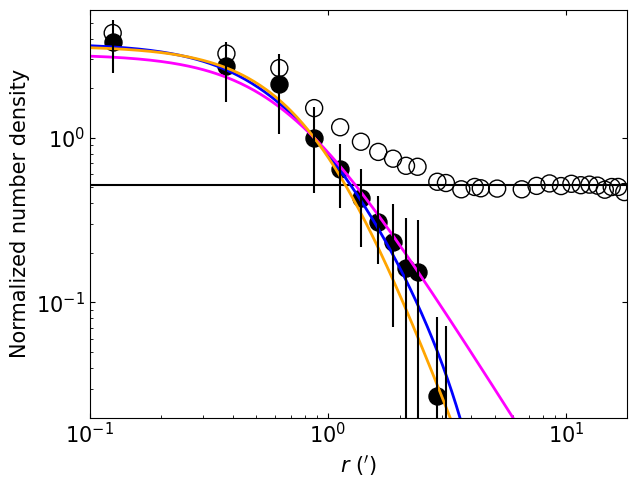}
\caption{Same as Figure~\ref{fig3} for HW~66}
\end{figure}

\begin{figure}
\includegraphics[width=\columnwidth]{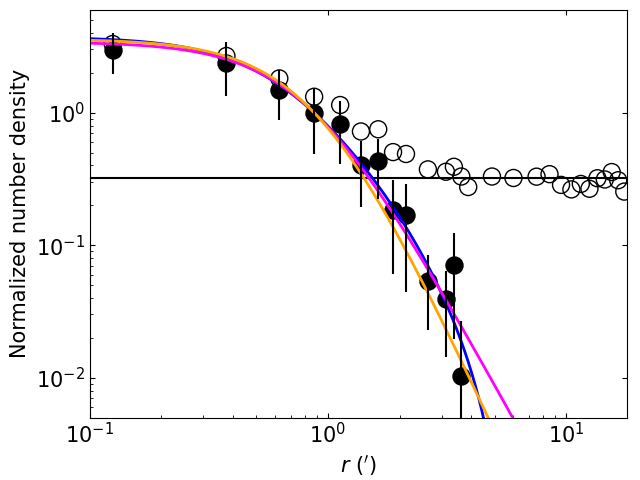}
\caption{Same as Figure~\ref{fig3} for HW~79}
\end{figure}

\begin{figure}
\includegraphics[width=\columnwidth]{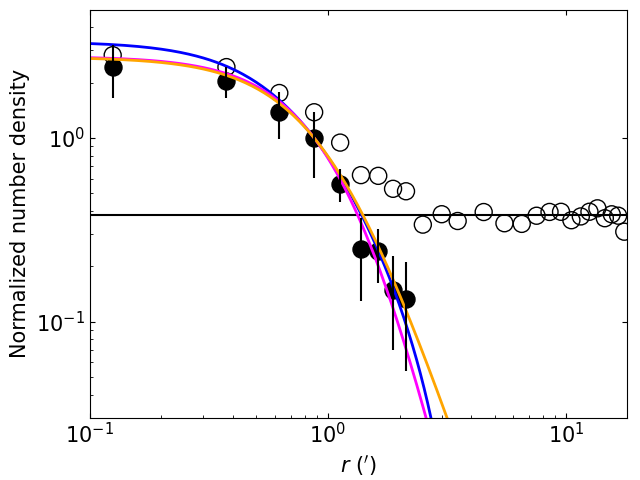}
\caption{Same as Figure~\ref{fig3} for K~1}
\end{figure}

\begin{figure}
\includegraphics[width=\columnwidth]{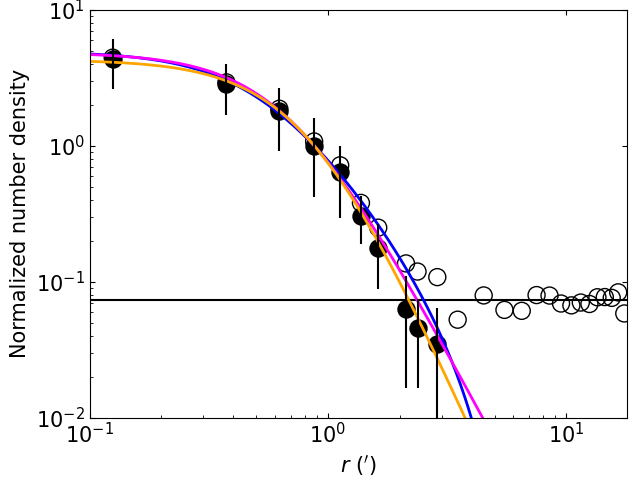}
\caption{Same as Figure~\ref{fig3} for K~3}
\end{figure}

\begin{figure}
\includegraphics[width=\columnwidth]{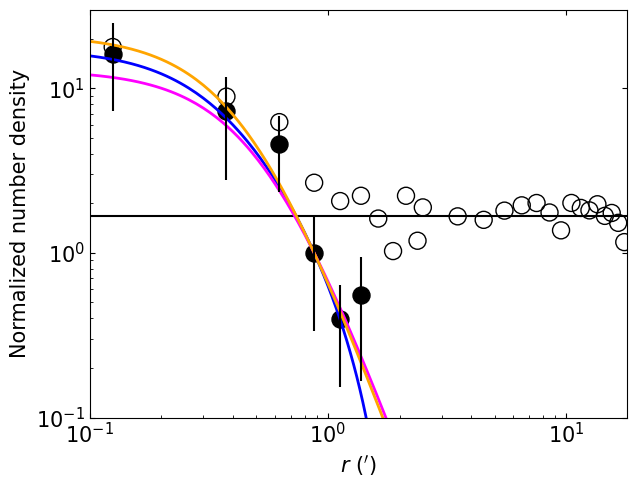}
\caption{Same as Figure~\ref{fig3} for L~2}
\end{figure}

\begin{figure}
\includegraphics[width=\columnwidth]{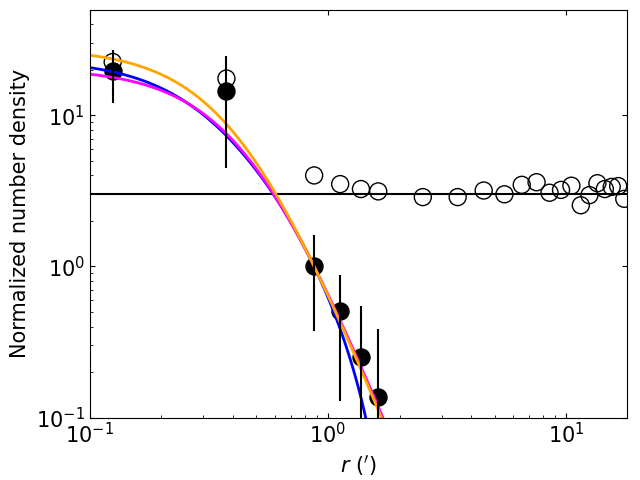}
\caption{Same as Figure~\ref{fig3} for L~109}
\end{figure}

\begin{figure}
\includegraphics[width=\columnwidth]{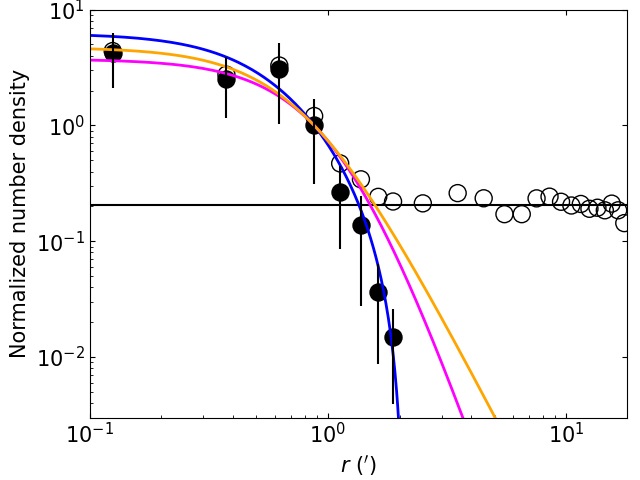}
\caption{Same as Figure~\ref{fig3} for L~110}
\end{figure}

\end{appendix}

\bsp	
\label{lastpage}
\end{document}